\documentclass[%
 prd, twocolumn, 
 amsmath,amssymb,
 aps, physrev,
]{revtex4-2}

\usepackage{graphicx}
\usepackage{dcolumn}
\usepackage{bm}

\usepackage{mathtools}
\usepackage{amsmath}
\usepackage{amssymb}
\usepackage{braket}
\usepackage{color}
\usepackage{xcolor}
\usepackage{hyperref}
\usepackage{slashed}
\usepackage{graphicx}
\usepackage{mathrsfs}
\usepackage{tikz}
\usepackage{tikz-cd}
\usepackage{enumitem}
\usepackage{eso-pic}

\usepackage{mathtools} 
\usepackage{extarrows} 
\usepackage{multirow}

\newcommand{\be}{\begin{equation}}
\newcommand{\ee}{\end{equation}}
\newcommand{\ba}{\begin{aligned}}
\newcommand{\ea}{\end{aligned}}
\newcommand{\bea}{\begin{eqnarray}}
\newcommand{\eea}{\end{eqnarray}}

\newcommand{\mD}{\mathcal{D}}

\newcommand{\reportnum}[2]{
  \AddToShipoutPictureBG*{%
    \AtPageUpperLeft{%
      \hspace{0.75\paperwidth}%
      \raisebox{#1\baselineskip}{%
        \makebox[0pt][l]{\textnormal{#2}}
  }}}%
}

\def\mb{\mathbb}

\def\mc{\mathcal}
\def\bp{\begin{pmatrix}}
\def\ep{\end{pmatrix}}

\def\tr{\mathop{\mathrm{tr}}\nolimits}

\newcommand{\newreptheorem}[2]{%
\newenvironment{rep#1}[1]{%
 \def\rep@title{#2 \ref{##1}}%
 \begin{rep@theorem}}%
 {\end{rep@theorem}}}
\makeatother

\newreptheorem{lemma}{Lemma}

\newreptheorem{conj}{Conjecture}

\newcommand{\Fbefore}{%
    \raisebox{-1.0ex}{\includegraphics[height=3.5ex]{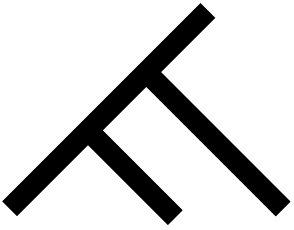}}%
}

\newcommand{\Fafter}{%
    \raisebox{-1.0ex}{\includegraphics[height=3.5ex]{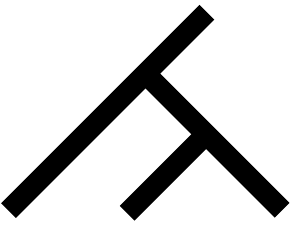}}%
}

\begin{document}

\reportnum{-3}{USTC-ICTS/PCFT-25-47}

\title{\textbf{Anomaly of Continuous Symmetries from Topological Defect Network} 
}%

\author{Qiang Jia$^1$}
\author{Ran Luo$^2$}
\author{Jiahua Tian$^3$}
\author{Yi-Nan Wang$^{2,4,5}$}
\author{Yi Zhang$^6$}

\affiliation{%
 $^1$Department of Physics, Korea Advanced Institute of Science \& Technology, Daejeon 34141, Korea
}
\affiliation{%
 $^2$School of Physics, Peking University 
}%
\affiliation{%
 $^3$School of Physics and Electronic Science, East China Normal University, Shanghai, China, 200241
}
\affiliation{$^4$Center for High Energy Physics, Peking University}

\affiliation{$^5$Peng Huanwu Center for Fundamental Theory, Hefei, Anhui 230026, China}

\affiliation{$^6$Kavli IPMU (WPI), UTIAS, The University of Tokyo, Kashiwa, Chiba 277-8583, Japan}


\date{\today}

\begin{abstract}
We show that the 't Hooft anomaly of a quantum field theory with continuous flavor symmetry can be detected from rearrangements of the topological defect webs implementing the global symmetry in general spacetime dimension, which is concretized in 2D by the \emph{F-moves} of the defect lines. Via dualizing the defects to flat background gauge field configurations, we characterize the 't Hooft anomaly by various cohomological data of the symmetry group, where the cohomology of Lie groups with discrete topology plays the central role. We find that an extra dimension emerges naturally as a consequence of the mathematical description of the 't Hooft anomaly in the case of flat gauging. 

\end{abstract}

\maketitle



\textit{Introduction.} A global symmetry of a Quantum Field Theory (QFT) is said to have a \emph{'t Hooft anomaly} if gauging it leads to an inconsistency, i.e. an obstruction to promoting it to a gauge symmetry~\cite{tHooft:1979rat}. The robustness of 't Hooft anomaly across ultraviolet to infrared makes it a particularly effective tool to study the properties of QFT even in the strongly coupled region~\cite{Witten:1983tw, Terning:1997xy}. The 't Hooft anomaly manifests itself as a nontrivial phase factor $e^{2\pi i \mathcal{A}(A,g)}$ acquired by $Z[A]$ under gauge transformation $A\rightarrow g^{-1}Ag + g^{-1}dg$, where $Z[A]$ is the paritition function of the QFT coupled to background field $A$ and for $\mathcal{A}(A,g)$ that cannot be cancelled by any local counterterm built out of $A$. Traditionally for continuous $G$, it can be evaluated using the \emph{descent equations} for small gauge transformations~\cite{Stora:1976kd,Stora:1983ct,Zumino:1983rz,Wess:1971yu,Callan:1984sa} and more generally shown to be the the \emph{eta-invariant} when $g$ is disconnected from the identity~\cite{Atiyah:1975jf,Witten:1982fp,Witten:1985xe,Witten:2019bou}.

Recent years have witnessed a revival of the study of 't Hooft anomalies in both condensed matter and high-energy physics \cite{Wen:2013oza,Kapustin:2013uxa,Freed:2014iua,Tachikawa:2017gyf,Bhardwaj:2017xup}. Thanks to the reformulation of the symmetry action as linking symmetry operator and charged defect in spacetime~\cite{Gaiotto:2014kfa,McGreevy:2022oyu,Schafer-Nameki:2023jdn,Bhardwaj:Lecture,Luo:2023ive,Shao:2023gho}, the action of finite group symmetries is now conveniently studied on equal footing with that of continuous symmetries. However, whereas the 't Hooft anomaly of discrete group symmetry can be computed directly as the \emph{F-symbol} associated with rearrangements of webs of defects, which are often described by elements of the group cohomology (e.g. $H^{d+1}(G,U(1))$ for a 0-form symmetry group $G$ in $d$-spacetime dimensions)\cite{Chen:2011pg,Kapustin:2014lwa,Kapustin:2014zva,Else:2014vma}, an analogous algorithm to compute the 't Hooft anomaly for continuous group symmetries \emph{in flat gauging} has long remained elusive.


In this work, we present such an algorithm for continuous symmetries, thereby filling the missing corner to place the treatment of finite and continuous symmetries on a completely parallel footing. We first show that a web of defects can be dualized to be a flat background field configuration. This effectively establishes the equivalence between the insertion of defects and coupling the conserved current to flat background fields. Via the dualization, we show in full generality how the 't Hooft anomaly manifests itself as the phase factor arising from rearrangements of the topological defects in the manner hinted in~\cite{Bhardwaj:Lecture}. We then compute the anomaly concretely via \emph{F-moves} of the \emph{topological defect lines} in 2D QFT~\cite{Moore:1988qv, Dijkgraaf:1989pz,Bhardwaj:2017xup,Tachikawa:2017gyf,Chang:2018iay} and show that the anomaly is characterized by cohomological data of $G$.

Very importantly, unlike the case of a finite group $G$ where the group cohomology $H^3(G,U(1))\cong H^3(BG,U(1))$ is unambiguously defined, one should be careful about the exact definition for Lie group cohomology. We emphasize that for a Lie group $G$ the correct version should be $H^3(G^\delta,U(1))$, which is defined to be the group cohomology when $G$ is equipped with discrete topology~\cite{Milnor1983_LieGroupsMadeDiscrete}. Despite that $H^3(G^\delta,U(1))$ is a highly complicated mathematical object, there exists an injection map from the group $H^4(BG,\mb{Z})$, where the anomaly polynomial for $G$ lives, to $H^3(G^\delta,U(1))$. Thus the usual anomaly polynomial $\sim F\wedge F$ can be elegantly embedded in this unified framework.

We find that the unifying mathematical structure underlying these various cohomology groups of $G$ suggests that the QFT with a 't Hooft anomaly naturally lives on the boundary of an \emph{anomaly theory} in one higher dimension. The generalization of the above cohomological characterization of 't Hooft anomaly to higher-dimensional spacetime is immediate in our formalism.

For detailed examples, we discuss the cases of a simple $G$ and $U(1)$ separately. For a compact simply connected and simple Lie group $G$, the $3d$ bulk anomaly theory is the non-abelian Chern--Simons invariant characterized by an integer $k\in \mathbb{Z}$ called the \emph{level}. It has a local expression as the  Chern--Simons 3-form $I_3=CS_k[A]=\frac{k}{2}\tr(A\wedge F-\frac{1}{3}A\wedge A\wedge A)$. We explicitly uplift the topological defect networks across the F-move to a three-dimensional bulk, and show that after assigning $F=0$, the cubic term in $CS_k[A]$ precisely generates the de Rham cohomology group $H_{\rm dR}^3(G,\mb R)$ for $G$.

For a non-simply connected Lie group such as $G=U(1)$, we point out that the field strength $F=dA$ would not vanish at the junction point for this case. This is not inconsistent with the requirement of flat gauging, as the precise definition of flat gauging should be the triviality of the holonomy $\exp\left(\oint_{\mc{C}}A\right)$ around a topologically trivial loop $\mc{C}$. Such a non-vanishing $F$ is precisely the source of the non-trivial $U(1)$ phase factor across the F-move, which we compute explicitly from the integration of $\epsilon F$ over the 2D spacetime, where $\epsilon$ is the finite gauge transformation parameter generating the F-move. The phase factor~\footnote{The same 3-cocycle expression was derived in~\cite{Kawagoe:2021gqi} for bosonic 2D theories using the method of Else and Nayak~\cite{Else:2014vma}. A recent discussion of $U(1)$ anomalies in the fermionic case within this framework can be found in~\cite{Okada:2025kie}.} 
\be
\label{U(1)-factor}
\omega(e^{i\alpha_1},e^{i\alpha_2},e^{i\alpha_3})=e^{i\alpha_1(\alpha_2+\alpha_3-[\alpha_2+\alpha_3])/(2\pi)}
\ee
with $\alpha_i \in [0, 2\pi)$ and $[r]$ denoting the part of $r$ exceeding $2\pi$, is consistent with the $N\rightarrow\infty$ limit of the generator of $H^3(\mb{Z}_N,U(1))$. 

We would also like to comment that (\ref{U(1)-factor}) is perfectly consistent with the newly proposed categorical formulation of continuous symmetry~\cite{Jia:2025vrj}. Namely, if one pretends that the definition of the symmetry category $\mathbf{Vec}_G^\omega$ and Drinfeld center $Z(\mathbf{Vec}_G^\omega)$ can be generalized to the case of $G=U(1)$, all the category theoretical data are consistent with the mathematical results in \cite{Freed:2009qp}, which are confirmed in \cite{Jia:2025vrj}.

Our results can be applied to all compact connected Lie groups. From the structure theorem~\cite{Borel1955TopologyLieGroups}, compact connected Lie groups are finitely covered by direct products of tori and simple non-abelian groups, such that their classification reduces to that of simple groups. The restriction to a compact connected and simple Lie group $G$ or $U(1)$ is sufficient in this context. 

Finally, although the detailed examples are discussed for the $d=2$ case, our framework applies to continuous 0-form symmetries in general $d$ spacetime dimensions as well.


\textit{Dualization of the defect web.} The VEV of a topological defect web $\mathcal{W}$ of a (Lagrangian) QFT $\mathcal{T}_G$ in $d$-dimensional spacetime $M_d$ can be computed as:
\begin{equation}
	\langle \mathcal{W} \rangle := \int \mD\psi\ e^{iS[\psi] + \int_{M_d} \mathrm{tr}(A_{\mathcal{W}}\wedge \star J)}\,,
\end{equation}
where $\psi$ stands for the set of local operators of $\mathcal{T}_G$ and $J(\psi)$ is the conserved current of the continuous global $G$-symmetry. $A_{\mathcal{W}}$ standards for the ordinary flat gauge field corresponding to the defect web $\mc{W}$. To illustrate the dualization, we consider a defect line $\mathcal{L}$ labeled by $g_\alpha = e^{i\alpha}\in G$ along $x = 0$ in $\mathbb{R}^2$, meaning that a particle charged under the global $G$-symmetry is transformed under a representation of $g_\alpha$ when crossing $x = 0$ from left to right, see (\ref{eq:Gauge_Convention}). We define the dualization $A_\mathcal{L}$ of $\mathcal{L}$ to be the background gauge field such that $\mathcal{P}e^{\int_\ell A_{\mathcal{L}}} = e^{i\alpha}$ for an arbitrary path $\ell$ crossing $x = 0$ once, which in turn means that $A_{\mathcal{L}}$ must be \emph{flat}. Given the condition, we can write:
\begin{equation}\label{eq:PD_of_TDL}
    A_{\mathcal{L}} = i\alpha \delta(x) dx = e^{-i\alpha H(x)} de^{i\alpha H(x)}
\end{equation}
where $H(x)$ is the Heaviside function, and we can indeed check that $dA_{\mathcal{L}} + A_{\mathcal{L}}\wedge A_{\mathcal{L}} = 0$. Equivalently, $\mathcal{L}$ defines the gauge configuration $A_{\mathcal{L}}|_{U_i} = g_i^{-1}dg_i$ for $g_0 = e$ and $g_1 = e^{i\alpha}$ on the following open cover of $\mathbb{R}^2$:
\begin{equation}\label{eq:Gauge_Convention}
    \begin{gathered}
	\begin{tikzpicture}
		\draw[dashed,->] (0,0) -- (3,0);
		\node[circle, fill=black, inner sep=1pt] at (1.5,0)  {};
		\node at (2,0)[above] {$x = 0$};
        \node at (1.5,1)[right] {$\mathcal{L}$};
		\draw[->] (1.5,1) -- (1.5,-1);
		\node at (0.5,0.7) {$U_0$};
		\node at (2.5,0.7) {$U_1$};
        \node at (3.3,0.2) {$A_{\mathcal{L}}$};
	\end{tikzpicture}
    \end{gathered}
\end{equation}
where the two patches $U_0 = (-\infty, \epsilon)$ and $U_1 = (-\epsilon,\infty)$ are glued along $(-\epsilon,\epsilon)$ for small positive $\epsilon$. Here we adopt the convention that the direction of the defect line rotates right-handedly with respect to the direction of the gauge transformation from $U_0$ to $U_1$. Clearly, the above discussions can be generalized to a codimension-one web of $\mathcal{W}$ in general $d$ dimensions.

The fundamental building block of any defect web (of lines) is the trivalent junction:
\begin{equation}
    \begin{tikzpicture}
        \draw[->] (-1,0) -- (-0.5,0);
        \draw[->] (-0.4,0) -- (-0.4,0.3);
        \draw[->] (0,0) -- (0.5,0);
        \draw (-0.5,0) -- (0,0);
        \draw[->] (0.7,0) -- (0.7,0.3);
        \draw[->] (0,0.5) -- (0.3,0.5);
        \draw (0.5,0) -- (1,0);
        \draw[->] (0,1) -- (0,0.6);
        \draw (0,0.6) -- (0,0);
        \node at (-1,0)[left] {$e^{i\beta}$};
        \node at (0,1)[right] {$e^{i\alpha}$};
        \node at (1,0)[right] {$e^{if(\alpha,\beta)}$};
        \node at (0,-0.3) {$(0,0)$};
        \node at (-0.4,0.4) {\scriptsize$\beta$};
        \node at (1,0.4) {\scriptsize$f(\alpha,\beta)$};
        \node at (0.4,0.6) {\scriptsize$\alpha$};
        
    \end{tikzpicture}
    \label{junction}
\end{equation}
The background gauge field $A_{\mc{L}}$ corresponding to the group element $e^{i\alpha}$ is also labeled, with $\alpha$. 
This trivalent junction can be smeared out to give $A = g^{-1}dg$ with $F=dA+A\wedge A=0$ for a globally defined $g$ on $\mathbb{R}^2$ for $G$ with trivial $\pi_1(G)$, since all paths from identity to $e^{i\alpha}e^{i\beta}$ on the group manifold are homotopic. However, a globally defined $g$ cannot be obtained by any kind of smearing for $G$ with non-trivial $\pi_1(G)$, e.g. for $G = U(1)$, and there exist flat configurations of $U(1)$ gauge fields with non-trivial $F=dA$. This subtlety is crucial for the derivation of anomalies for abelian groups.

Now we clarify the notion of flat gauging of continuous $G$ global symmetry, given by the following two equivalent conditions:
\begin{enumerate}
\item A flat gauging is equivalent to summing over all possible insertions of topological defect networks in the new partition function.
\item Summing over configurations of gauge fields $A$, where the holonomy of the gauge field around a topologically trivial loop $\mathcal{P} \exp (\oint_\mathcal{C} A) = e$, the identity element of $G$.
\end{enumerate}


\textit{'t Hooft anomaly from F-move and the Lie group cohomology.} Having defined the dualization for continuous $G$, one can compare $\langle \mathcal{W} \rangle$ with $\langle \mathcal{W}' \rangle$ by coupling $\mathcal{T}_G$ to flat $A_{\mathcal{W}}$- and $A_{\mathcal{W}'}$-backgrounds, respectively. Suppose $A_{\mathcal{W}'} = e^{-i\Lambda}A_{\mathcal{W}}e^{i\Lambda} + e^{-i\Lambda}de^{i\Lambda}$ for gauge transformation $\psi\rightarrow e^{i\Lambda} \psi$ parametrized by $\Lambda$, we have:
\begin{equation}\label{eq:Anomaly_Web}
	\langle \mathcal{W}' \rangle = \int \mathcal{D} \psi\ e^{iS[\psi] +\int_{M_d} \mathrm{tr}(A_{\mathcal{W}'}\wedge \star J)} = e^{2\pi i\mathcal{A}[A_{\mathcal{W}};\Lambda]} \langle \mathcal{W} \rangle 
\end{equation}
where $\mathcal{A}[A_{\mathcal{W}};\Lambda] \neq 0$ is the quantum anomaly. 
Therefore, given our dualization, the computation of anomaly arises from rearrangements of the defect webs amounts to finding the gauge transformation interpolating the dual flat field configurations.

As a reminder, the 't Hooft anomaly of $\mathbb{Z}_m$-symmetry of a 2D bosonic theory is well-known to be characterized by $H^3(\mathbb{Z}_m, U(1))$, the 3rd group cohomology of $\mathbb{Z}_m$ with $U(1)$-coefficient, in which the phase factor $e^{2 \pi i a_1(a_2+a_3 - \overline{a_2+a_3})/(m) }$ of the F-move of the defect webs in Figure~\ref{fig:F-move} lives~\cite{Moore:1988qv}. 
\begin{figure}[h]
    \centering
    \begin{tikzpicture}
		\draw (0,0) -- (1.5,1.5);
        \draw[->] (0,0) -- (0.2,0.2);
        \draw[->] (0.5,0.5) -- (0.7,0.7);
        \draw[->] (1,1) -- (1.2,1.2);
		\draw (0.5,0.5) -- (1,0);
        \draw[->] (1,0) -- (0.7,0.3);
		\draw (1,1) -- (2,0);
        \draw[->] (2,0) -- (1.5,0.5);
		\node at (0,-0.2) {$e^{2 \pi i \frac{a_1}{m}}$};
		\node at (1,-0.2) {$e^{2 \pi i \frac{a_2}{m}}$};
		\node at (2,-0.2) {$e^{2 \pi i \frac{a_3}{m}}$};
		\node at (1.75, 0.75) {$=$};
		\node at (3.8,0.85) {$e^{2 \pi i \frac{a_1(a_2+a_3 - \overline{a_2+a_3})}{m} }\times$};
		\draw (5,0) -- (6.5,1.5);
        \draw[->] (5,0) -- (5.7,0.7);
        \draw[->] (6,1) -- (6.2,1.2);
		\draw (6.5,0.5) -- (6,0);
        \draw[->] (6.5,0.5) -- (6.2,0.8);
        \draw[->] (6,0) -- (6.2,0.2);
        \draw[->] (7,0) -- (6.8,0.2);
		\draw (6,1) -- (7,0);
		\node at (5,-0.2) {$e^{2 \pi i \frac{a_1}{m}}$};
		\node at (6,-0.2) {$e^{2 \pi i \frac{a_2}{m}}$};
		\node at (7,-0.2) {$e^{2 \pi i \frac{a_3}{m}}$};
	\end{tikzpicture}
    \caption{The F-move of defect web for $G=\mb{Z}_m$ 0-form symmetry group in 2D. $a_i \in \{0,1,\ldots,m-1\}$ and $\overline{m}$ is defined as $m$ mod $\mb{Z}$.}
    \label{fig:F-move}
\end{figure}
In $d$-spacetime dimensions the characterization is given by $H^{d+1}(G, U(1))$ for discrete $G$~\cite{Kapustin:2014zva, Tachikawa:2017gyf}. To fill the gap of a similar characterization for continuous $G$, we will focus on continuous $G$ of interests, and compute their anomalies in the case of \emph{flat gauging}, i.e. from coupling $\mathcal{T}_G$ to \emph{flat} background field that is the closest cousin of the discrete case where all background fields are automatically flat.

We show that the 't Hooft anomaly is encoded in the phase factor in $\langle \Fbefore \rangle = e^{2\pi i\mathcal{A}[A_{\scalebox{0.3}{\Fbefore}};\Lambda]} \langle \Fafter \rangle$ (cf.~(\ref{eq:Anomaly_Web})) for the defect webs $\Fbefore$ and $\Fafter$ living on $M_2$ in $\mathcal{T}_G$ with Lie group $G$. Here $\Lambda$ is the gauge transformation from $A_{\scalebox{0.3}{\Fbefore}}$ to $A_{\scalebox{0.3}{\Fafter}}$.

 In the modern langauge~\cite{Witten:2019bou}, $\mathcal{A}[A_{\scalebox{0.3}{\Fbefore}};\Lambda]$ can be calculated as (cf. Eq. \eqref{sm-anomaly} in Appendix~\ref{sec:anomaly-gauge-trans}):
\begin{equation}\label{eq:FiniteParam_Anomaly}
    \mathcal{A}(A_{\scalebox{0.3}{\Fbefore}};\Lambda) = \int_{M_3} CS_k[A(t)]\,,
\end{equation}
using the level-$k$ Chern--Simons invariant $CS_k$ where $t\in[0,1]$ parametrizing a mapping cylinder in the space of gauge field configurations along which $A(t) := g(t)^{-1}A_{\scalebox{0.3}{\Fbefore}} g(t) + g(t)^{-1} d g(t)$ for $g(x,y,t=0) = 1\in G$ and $g(x,y,t=1) = e^{i\Lambda(x,y)}$ interpolates from $A_{\scalebox{0.3}{\Fbefore}}$ to $A_{\scalebox{0.3}{\Fafter}}$. We use the normalization $CS_k[A] := \frac{k}{2} \tr(A \wedge dA + \frac{2}{3} A\wedge A \wedge A ) = \frac{k}{2} \tr(A \wedge F - \frac{1}{3} A\wedge A \wedge A )$ on a local patch of $M_3$. 
For simplicity, we restrict to theories whose Chern--Simons invariants used in anomaly computations originate from integral classes in $H^4(BG,\mathbb{Z})$ (cf. Eq.~\eqref{eq:cano2} in Appendix~\ref{sec:anomaly-coh}). This condition constrains the allowed minimal level $k$, which depends on $G$. We do not discuss these cases individually and omit the subscript $k$ for brevity~\footnote{For some compact Lie groups, such as $U(1)$ and $SO(3)$, defining the Chern–Simons invariant at level $k=1$ requires a spin structure~\cite{Dijkgraaf:1989pz,Belov:2005ze}. In these cases, one replaces the Chern–Simons term by the corresponding $\eta$-invariant (or equivalently, a quadratic refinement of the gauge fields pairing). This corresponds to \emph{one half} of the generator of $H^4(BG,\mathbb{Z})$ and therefore can not be directly obtained as a class in $H^4(BG,\mathbb{Z})$. In this letter, we only require the orientability of the spacetime manifold and its extension throughout.}.

The interpolation from $A_{\scalebox{0.3}{\Fbefore}}$ to $A_{\scalebox{0.3}{\Fafter}}$ is illustrated in the leftmost graph in Figure~\ref{fig:Deformations}. It is not hard to see that by our convention, the configuration in the front is equivalent to $\Fbefore$ and the one in the back is equivalent to $\Fafter$. Since $\mathcal{A}[A_{\scalebox{0.3}{\Fbefore}};\Lambda]$ is clearly independent from small deformations of the mapping cylinder, i.e. the details of $g$, as long as boundary configurations $A_{\scalebox{0.3}{\Fbefore}}$ and $A_{\scalebox{0.3}{\Fafter}}$ are fixed, we further glue the vertices of $\Fafter$ with the ones with the same labels in $\Fbefore$ at the two copies $M_2\times \{0\}$ and $M_2\times \{1\}$ to arrive at the configuration in the middle of Figure~\ref{fig:Deformations}. This is the difference between the corresponding webs of defects and the middle configuration can be further deformed to be a decorated tetrahedron as in the rightmost of Figure~\ref{fig:Deformations}, where the directions of the defects and of the gauge transformations are set by our convention~(\ref{eq:Gauge_Convention}). This process can be understood as the anomaly-inflow---very similar to the finite group case~\cite{Tachikawa:2017gyf}---i.e. an extra 3-simplex can be attached in the bulk to cancel the anomaly and the decoration of the tetrahedron in the bulk describes the pullback of the 3-simplex $\langle e, g_1, (g_1g_2), (g_1g_2g_3)\rangle$ in $BG^\delta$ to the bulk. Here $G^\delta$ is the Lie group $G$ with discrete topology and $BG^\delta$ is the \emph{Eilenberg--MacLane} space $K(G,1)$ classifying flat $G$-bundles, we denote the classifying map by $\varphi^\delta:M_3\rightarrow BG^\delta$~\footnote{See more details on $BG^\delta$ in Appendix~\ref{sec:anomaly-coh}. For a discussion of the simplicial complex construction of $BG$ for finite $G$, the reader may consult section 9.1 in the lecture note \url{https://member.ipmu.jp/yuji.tachikawa/lectures/2024-mathphys/notes.pdf}.}.

 Similar to the finite group case, in terms of the classifying space, the anomaly is obtained by evaluating a cohomology class $[\Omega] \in H^3(BG^\delta,\mathbb{R}/\mathbb{Z})$ on $\langle e, g_1, (g_1g_2), (g_1g_2g_3)\rangle$. Here we treat the coefficient $U(1) = \mathbb{R}/\mathbb{Z}$ as an additive group. Pullback with $\Omega$ with $\varphi^\delta$  and integrate over the 3D spacetime, we obtain the Dijkgraaf-Witten phase
\begin{equation} \label{eq:Omega}
     \mathcal{A}(A_{\scalebox{0.3}{\Fbefore}};\Lambda) = \int_{M_3} (\varphi^\delta)^*(\Omega) \,,
\end{equation}
in other words, we claim that $\Omega$ found by F-move pulls back to the Chern--Simons invariant $CS$. See the proof of \eqref{eq:Omega} in Appendix~\ref{sec:anomaly-coh}.

\begin{widetext}
    \begin{center}
        \begin{figure}[htbp]
            \includegraphics[width=1.0\textwidth]{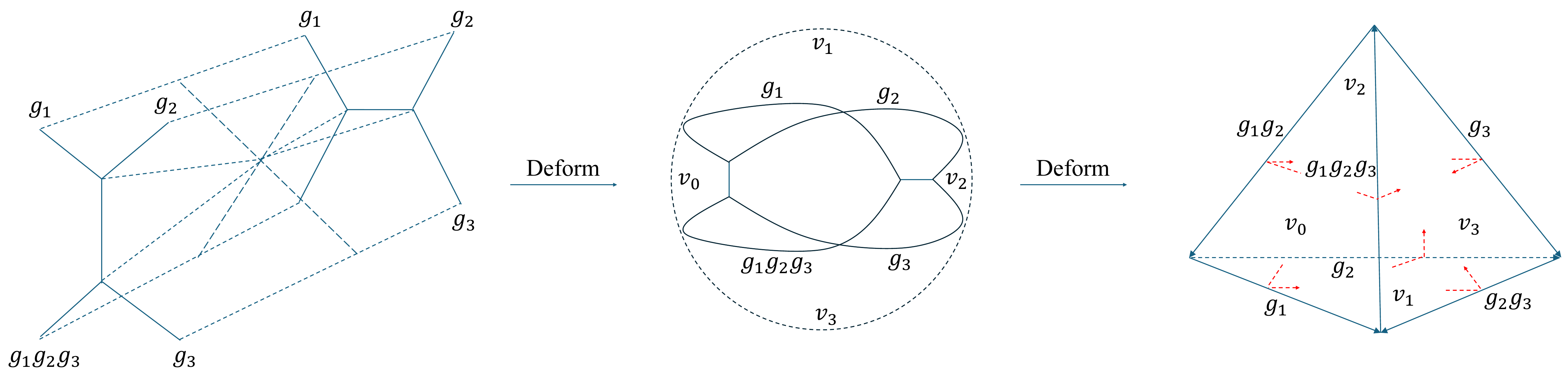}
            \caption{The deformations of the combined $\Fbefore$ and $\Fafter$ configuration. The group multiplications in this figure are all taken from the right.}
            \label{fig:Deformations}
        \end{figure}
    \end{center}
\end{widetext}

By investigating the local expression of Chern--Simons invariant, we can get some other manifestations of anomalies. For this discussion, as $G$ is simply connected we can assign $F=0$ for flat gauge field.
In the $M_3$ bulk, our gauge field $A(t)$ is flat. On a local patch $U_i$, we have $A_i:=A|_{U_i} = \tilde g_i^{-1}d \tilde g_i$ for some $\tilde g_i:U_i\rightarrow G$ and 
\begin{equation}\label{eq:AAA_to_omega3}
    \int_{U_i} - \frac{k}{6} \tr( A_i \wedge A_i\wedge A_i) = \int_{U_i} k \,\tilde  g_i^* \omega_3 = \int_{\tilde g_i(U_i)} k \,\omega_3
\end{equation}
where $\tilde g_i^*$ is the pull-back from $\Omega^*(G)$ to $\Omega^*(U_i)$ and $\tilde g_i^*\omega_3 = \frac{i^2}{3!} \tr(\tilde g_i^{-1}d \tilde g_i)^3$ with $\omega_3$ being a \emph{left-invariant 3-form} on $G$. For a compact, connected, simply connected and simple $G$, $\omega_3$ generates the de Rham cohomology $H_{\rm dR}^3(G,\mathbb{R}) \cong \mathbb{R}$, it is also the image of an integral class in the singular cohomology $H_{\rm singular}^3(G,\mathbb{Z})\cong \mathbb{Z}$~\footnote{We use the convention $\omega_{2n-1}:=i^n \frac{(n-1)!}{(2n-1)!} \tr(\tilde g^{-1}d \tilde g)^{2n-1}$, such that it is normalized to integrate to 1 on the generator of $\pi_{2n-1}(G)$, and for $n=2$, the generator of $\pi_{3}(G)$ is sent to the generator of $H_{3}(G,\mathbb{Z})$~\cite{Lee:2020ojw}}. Furthermore, such $G$-bundle on any 3-manifold is necessarily trivial. Hence $\tilde g_i = \tilde g$ is globally defined, and the anomaly can be calculated as:
\begin{equation}\label{eq:Anomaly_to_omega3}
    \mathcal{A}(A_{\scalebox{0.3}{\Fbefore}};\Lambda) = \int_{ \tilde g(M_3)} k \,\omega_3\,.
\end{equation}

The right-hand side of \eqref{eq:Anomaly_to_omega3} now has another interpretation, namely, the action of ungauged 2D \emph{Wess--Zumino--Witten} model. Indeed, this is a fact of anomaly matching~\cite{Wess:1971yu,Novikov:1982ei,Witten:1983tw}.
The appearance of the left invariant 3-form $\omega_3$ as a \emph{suspension} \eqref{eq:susp} (the inverse is referred to as \emph{transgression}~\cite{Borel1955TopologyLieGroups,Dijkgraaf:1989pz}) of $CS$ on the fiber $G$ via the bundle fibration $G\rightarrow P \rightarrow M_3$ was discussed originally in~\cite{Chern:1974ft}, see also discussions~\cite{Dijkgraaf:1989pz,Witten:1991mm, Freed:2006mx,Lee:2020ojw} related to anomalies.

On the other hand, we take $\langle X,Y \rangle = \tr(XY) $ to be the Killing form of the Lie algebra $\mathfrak{g}$ of $G$ then 
\begin{equation}
  -\frac{k}{6}  \tr( A \wedge A\wedge A) =  -\frac{k}{12}\langle A, [A, A] \rangle \,,
\end{equation}
where we recognize that $\langle \cdot, [\cdot,\cdot] \rangle$ is the generator of the $3^{\text{rd}}$ \emph{Lie algebra cohomology} $H^3(\mathfrak{g},\mathbb{R})$ of the flavor algebra $\mathfrak{g}$. 
We see that the group $H^3(\mathfrak{g},\mathbb{R})$ also classifies the local density of anomaly in flat gauging. 

Since our $G$ is taken to be compact, connected and simple, we always have $H^3(\mathfrak{g},\mathbb{R}) \cong H_{\rm dR}^3(G,\mathbb{R}) \cong \mathbb{R}$. As the 3-forms have integral periods, they are the image of the lattice $\mathbb{Z} \cong H_{\rm dR}^3(G,\mathbb{Z})$ inside $H_{\rm dR}^3(G,\mathbb{R})$. The coefficient exchanging map is in fact injective~\cite{Baez:2003yaq}
\begin{equation}\label{eq:HdR=Hg}
    H_{\rm singular}^3(G,\mathbb{Z}) \cong H_{\rm dR}^3(G,\mathbb{Z}) \hookrightarrow H_{\rm dR}^3(G,\mathbb{R}) \cong H^3(\mathfrak{g},\mathbb{R})\,.
\end{equation}
Recall that the suspension map \eqref{eq:susp} 
\begin{equation} 
    \tau : H^4(BG,\mathbb{Z}) \longrightarrow H^{3}_{\rm singular}(G,\mathbb{Z}) 
\end{equation}
is also an isomorphism if $G$ is in addition simply connected. For these groups, we have \begin{equation}
     H_{\rm singular}^3(G,\mathbb{Z}) \cong H^4(BG,\mathbb{Z}) \xhookrightarrow{\kappa^\delta} H^3(BG^\delta,\mathbb{R}/\mathbb{Z}) \,,
\end{equation}
where the last map \eqref{eq:cano1} is also an injection~\cite{Baez:2003yaq}.

\textit{The emergence of extra dimension.} Moreover, there is a homomorphism $w: H^3(\mathfrak{g},\mathbb{R})\rightarrow H^3(B\overline{G},\mathbb{R})$ where $B\overline{G}$ is the \emph{classifying space of flat $G$-trivial bundles}~\cite{morita2001geometry}. One particularly interesting feature about $B\overline{G}$ is that a trivialization can be explicitly written down as an element of $\overline{G} = \{(g,\ell) \in G^\delta\times G^I | \ell(0) = g, \ell(1) = e \}$ where $G^\delta$ is $G$ equipped with discrete topology together with the canonical map $\iota:G^\delta \rightarrow G$ and $G^I := \text{Map}(I, G)$ for $I := [0,1]$.

More precisely, we consider a web of defects on a general 2D spacetime manifold $M_2$ which, by Poincar\'e duality, defines a flat $G$-bundle over $M_2$ determined up to isomorphism by the homotopy class of the \emph{classifying map} $f^\delta: M_2\rightarrow BG^\delta$ and a trivial bundle given by a null-homotopic map $f^n:M_2\rightarrow BG$. Using the lift of $\iota$, $B\iota : BG^\delta\rightarrow BG$, the map $\overline{f} : (x,t) \mapsto (f^\delta(x), \gamma_x(t)) \in BG^\delta\times BG$ where $\gamma_x(0) = B\iota(f^\delta(x))$ defines a homotopy from $f^\delta$ to $f^n$ by the \emph{mapping path space construction}~\cite{may1999concise}. Equivalently, $\overline{f}$ is a map from $M_2\times [0,1]$ to $B\overline{G} \subset BG^\delta\times BG$. Therefore, we have:
\begin{equation}\label{eq:BGbar_to_M2I}
    \int_{\overline{f}(M_2\times I)} w(\omega^3) = \int_{M_2\times I} \overline{f}^*\circ w (\omega^3)
\end{equation}
where $\overline{f}^*$ is the pull-back of $\overline{f}$.

\begin{figure}[h]
    \centering
    \includegraphics[height=3.5cm]{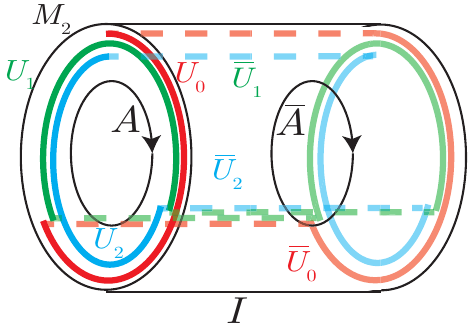}
    \caption{The showing of open cover $U_k$ $(k=0,1,2)$ of $M_2$ as well as the open cover $\overline{U}_k$ $(k=0,1,2)$ of $M_2\times I$.}
    \label{fig:patches}
\end{figure}

To investigate the abstract map $\overline{f}^*\circ w$ at a more elementary level, we fix $\pi_1(M_2) = \mathbb{Z}$ for simplicity (e.g. let $M_2\cong S^1\times \mathbb{R}$). We use the open cover $U_k = (\frac{2k\pi}{3}, \frac{(2k+4)\pi}{3})$ (modulo $2\pi$) of $S^1$, $k=0,1,2$, and define the flat connection $A = \frac{i\alpha}{2\pi}d\theta$ $(\alpha\in i\mathfrak{g})$ on a trivial bundle with the transition function $g_{ij} \equiv 1$ on $U_{ij} := U_i\cap U_j$ from $U_i$ to $U_j$. We plot the spacetime geometry and patches in Figure~\ref{fig:patches}.  Under gauge transformation $u_k = e^{-\frac{i\alpha}{2\pi} (\theta_k + \frac{2k\pi}{3})}$ on each $U_k$ parametrized by $\theta_k\in (0,\frac{4}{3}\pi)$, $A_k$ vanishes because $u_k^{-1}A_k u_k + u_k^{-1} du_k = 0$. Meanwhile, the transition function on $U_{i,i+1}$ becomes $g'_{i,i+1} = u_i^{-1} g_{i,i+1} u_{i+1} = u_i^{-1}u_{i+1}$. Since $\theta_i(p) = \theta_{i+1}(p) + \frac{2\pi}{3}$, $\forall p\in U_{i,i+1}$ for $i$ mod 3, we have:
\begin{equation}\label{eq:nontrivial_trans}
	g'_{01}(U_{01}) = g'_{12}(U_{12}) = 1,\ g'_{20}(U_{20}) = e^{i\alpha} \,. 
\end{equation}
Hence, while $u_k$ trivializes $A$, it yields non-trivial transition functions parametrized by $\alpha$. To uplift $A$ to $M_2\times I$, we define $\overline{u}_k = e^{-\frac{i\alpha}{2\pi} (\theta_k + \frac{2k\pi}{3})f(t)}$ with $f(0) = 1$ and $f(1) = 0$ for $t\in I$ as a function from $\overline{U}_k := U_k\times \mathbb{R} \times I$ to $G$. We further consider the gauge field $\overline{A}_k = \overline{u}_k^{-1}A_k\overline{u}_k + \overline{u}_k^{-1} d\overline{u}_k$ on $\overline{U}_k$ and the transition function $\overline{g}_{ij} = \overline{u}_i^{-1}\overline{u}_j$ on $\overline{U}_{ij}$. The pair $(\overline{g}_{ij}, \overline{A}_i)$, defined on (an open cover of
) $M_2\times I$, leads to a trivialization of a bundle with non-trivial transition function~(\ref{eq:nontrivial_trans}) over $M_2\times \{0\}$ to a trivial bundle with non-zero flat connection over $M_2\times \{1\}$. In this example, $\overline{g}_{20} = e^{i\alpha f(t)}$ provides a concrete physical realization of an element $(e^{i\alpha}, \ell)$ of $\overline{G}$.

Since $\overline{A}_k = \overline{g}_k^{-1}d\overline{g}_k$ with  $\overline{g}_k = u_k^{-1}\overline{u}_k$ is flat, we can replace $(A,U)$ in~(\ref{eq:AAA_to_omega3}) by $(\overline{A}_k,\overline{U}_k)$. Together with~(\ref{eq:BGbar_to_M2I}) restricted to $\overline{U}_k$ and a suitable choice of normalization, we are led to~\footnote{Actually, one can define $w$, which has not yet been specified explicitly, to be the homomorphism from $H^3(\mathfrak{g},\mathbb{R})$ to $H^3(B\overline{G},\mathbb{R})$ such that $\overline{f}^*\circ (k w|_{\overline{U}_k}) = \omega_3|_{\overline{U}_k}$ holds on each local patch $\overline{U}_k$.}:
\begin{equation}
    \overline{f}^*\circ w (\omega_3)|_{\overline{U}_k} = \overline{g}^*(\omega_3)|_{\overline{U}_k} = CS[\overline{A}]|_{\overline{U}_k}\,.
\end{equation}
Physically, this means that the data of 't Hooft anomaly can be captured by the Chern-Simons action of a flat gauge field on $M_2\times I$ with non-trivial transition functions at $M_2\times\{0\}$ which trivializes at $M_2\times \{1\}$ at the price of yielding non-zero flat connection. Moreover, the characterization of 't Hooft anomaly by $H^3(\mathfrak{g},\mathbb{R})$ naturally requires the emergence of an extra dimension of the anomaly theory, manifests itself as $I$, because of the structure of $\overline{f}^*\circ w$.

\textit{$U(1)$ flavor symmetry and its 't Hooft anomaly.} A mathematically cautious reader might have already noticed that the previous calculation leading to $H^3(G,\mathbb{R})$ fails for $U(1)$, since both $H^3(U(1),\mathbb{R})$ and $H^3(\mathfrak{u}(1),\mathbb{R})$ vanish. Moreover, the anomaly polynomial for $U(1)$ that generates $H^4(BU(1),\mb{Z})\cong\mb{Z}$ is $CS[A]=A\wedge dA$, where the cubic term $\frac{1}{3}A\wedge A\wedge A$ is absent. To get a non-zero anomaly, we comment that there is a non-zero field strength at the junction point $(x,y)=(0,0)$ in (\ref{junction}).

For the case of $G=U(1)$, we have $f(\alpha,\beta)=(\alpha+\beta)\ \mathrm{mod}\ 2\pi\mb{Z}\equiv[\alpha+\beta]$. The flat gauge field dual to the trivalent junction is then
\be
A=i\alpha\delta(x)H(y)dx+i(\beta H(-x)+[\alpha+\beta]H(x))\delta(y) dy\,,
\ee
and the corresponding field strength
\be
F=dA=i([\alpha+\beta]-\alpha-\beta)\delta(x)\delta(y)dx\wedge dy
\ee
is non-vanishing at the junction point.

Now to compute the 't Hooft anomaly of $U(1)$ flavor symmetry, we apply the previous computation to the configuration in Figure~\ref{fig:Fmove_U1}.
\begin{widetext}
\begin{center}
    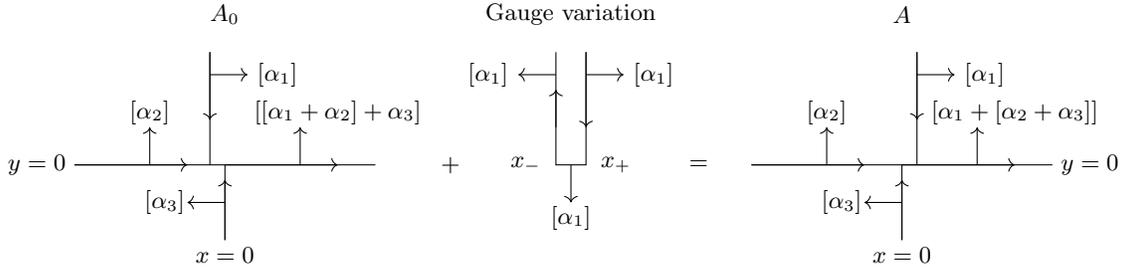
\begin{figure}[h]
	\begin{tikzpicture}
		\draw (-2,0) -- (2,0);
        \draw[->] (-2,0) -- (-0.5,0);
        \draw[->] (0,0) -- (1.5,0);
		\draw (0,-1) -- (0,0);
		\draw[->] (0,-0.5) -- (-0.5,-0.5);
		\draw[->] (-1,0) -- (-1,0.5);
		\draw[->] (1,0) -- (1,0.5);
		\draw (-0.2,0) -- (-0.2,1.5);
        \draw[->] (-0.2,1.5) -- (-0.2,0.6);
		\draw[->] (-0.2,1.2) -- (0.3,1.2);
        \draw[->] (0,-1) -- (0,-0.2);
		\node at (-0.8,-0.5) {$[\alpha_3]$};
		\node at (-1,0.7) {$[\alpha_2]$};
		\node at (1.5,0.7) {$[[\alpha_1 + \alpha_2] + \alpha_3]$};
		\node at (0.7,1.2) {$[\alpha_1]$};
		\node at (-2.5,0) {$y = 0$};
		\node at (0,-1.2) {$x = 0$};
		\node at (0,2.0) {$A_0$};
		
		\node at (3.0,0) {$+$};

        \draw[->] (4.4,0.5) -- (4.4,1);
        \draw[->] (4.8,1.5) -- (4.8,0.5);
		\draw (4.+0.4,1.5) -- (4.+0.4,0) -- (4.4+0.4,0) -- (4.4+0.4,1.5);
		\draw[->] (4+0.4,1.2) -- (3.5+0.4,1.2);
		\draw[->] (4.4+0.4,1.2) -- (4.9+0.4,1.2);
		\draw[->] (4.2+0.4,0) -- (4.2+0.4,-0.5);
		\node at (4.2+0.4,-0.7) {$[\alpha_1]$};
		\node at (3.5,1.2) {$[\alpha_1]$};
		\node at (5.7,1.2) {$[\alpha_1]$};
		\node at (3.6+0.4,0) {$x_-$};
		\node at (4.8+0.4,0) {$x_+$};
		\node at (4.6,2.0) {Gauge variation};
		
		\node at (6.3,0) {$=$};

        \draw[->] (-2+9,0) -- (-0.5+9,0);
        \draw[->] (0+9,0) -- (1.5+9,0);
		\draw (-2+9,0) -- (2+9,0);
		\draw (0+9,-1) -- (0+9,0);
        \draw[->] (0+9,-1) -- (0+9,-0.2);
		\draw[->] (0+9,-0.5) -- (-0.5+9,-0.5);
		\draw[->] (-1+9,0) -- (-1+9,0.5);
		\draw[->] (1+9,0) -- (1+9,0.5);
		\draw (-0.2+9.4,0) -- (-0.2+9.4,1.5);
        \draw[->] (-0.2+9.4,1.5) -- (-0.2+9.4,0.6);
		\draw[->] (-0.2+9.4,1.2) -- (0.3+9.4,1.2);
		\node at (-0.8+9,-0.5) {$[\alpha_3]$};
		\node at (-1+9,0.7) {$[\alpha_2]$};
		\node at (1.5+9,0.7) {$[\alpha_1 + [\alpha_2 + \alpha_3]]$};
		\node at (0.7+9.4,1.2) {$[\alpha_1]$};
		\node at (-2.5+14,0) {$y = 0$};
		\node at (0+9,-1.2) {$x = 0$};
		\node at (0+9,2.0) {$A$};
	\end{tikzpicture}
	\caption{The F-move of defect lines in a theory with $U(1)$ flavor symmetry. A topological line carrying $e^{i\alpha}\in U(1)$ is labeled by $[\alpha] := \alpha \mod \ 2\pi\mathbb{Z}$, with an arrow labeling the direction of gauge transformation in the same sense as in (\ref{junction}).}\label{fig:Fmove_U1}
\end{figure}
\end{center}
\end{widetext}

It is easy to see that the left-most and the right-most configurations in Figure~\ref{fig:Fmove_U1} are topologically equivalent to $\Fbefore$ and $\Fafter$, respectively. Hence our task is to find their Poincar\'e duals $A_{\scalebox{0.3}{\Fbefore}}$ and $A_{\scalebox{0.3}{\Fafter}}$, and the gauge transformation parameter $d\epsilon \equiv id\Lambda= A_{\scalebox{0.3}{\Fafter}} - A_{\scalebox{0.3}{\Fbefore}}$. Actually, using the configuration in the middle of Figure~\ref{fig:Fmove_U1} and assuming $\alpha <2\pi$ hence $[\alpha_1] = \alpha_1$, one can show that (see the details in Appendix~\ref{sec:anomaly-U(1)}):
\begin{equation}
    \epsilon = - i\alpha_1 (H(x_-) - H(x_+))H(y)
\end{equation}
This enables us to compute the 't Hooft anomaly to be
\begin{equation}
    \mathcal{A}[A_{\scalebox{0.3}{\Fbefore}};\epsilon] = \frac{1}{(2\pi)^2}\int_{\mathbb{R}^2} \epsilon dA_{\scalebox{0.3}{\Fbefore}} = \frac{1}{(2\pi)^2}\alpha_1 (\alpha_2 + \alpha_3 - [\alpha_2 + \alpha_3])\,.
\end{equation}

Viewing 
\be
\ba
f(e^{i\alpha_1}, e^{i\alpha_2}, e^{i\alpha_3})& := e^{2\pi i \mathcal{A}[A_{\scalebox{0.3}{\Fbefore}};\epsilon]}\cr
&=e^{i\alpha_1 (\alpha_2 + \alpha_3 - [\alpha_2 + \alpha_3])/(2\pi)}
\label{U(1)-f}
\ea
\ee
as a map $f: G^3\rightarrow U(1)$, one can check that $\delta f = 0$ where $\delta$ is the coboundary operator for the Lie group cohomology of $U(1)^\delta$ (see an explicit computation in the appendix~\ref{sec:anomaly-U(1)}). 

The map (\ref{U(1)-f}) is obviously not continuous, which aligns with the usage of group cohomology $H^3(U(1)^\delta, U(1))$ with discrete topology. 
Comparing to the phase factor in Figure~\ref{fig:F-move}, we see that (\ref{U(1)-f}) is exactly the $m\rightarrow\infty$ limit of the generator of $H^3(\mb{Z}_m,U(1))$. One can further show that this map defines a non-trivial algebraic 3-cocycle~\cite{Kawagoe:2021gqi}.

At the group level, the direct limit of $\mathbb{Z}_m$ is $\mathbb{Q}/\mathbb{Z}$, the union of all roots of unity, as the torsion subgroup of $U(1)^\delta$. It is well-known that $H^3(\mathbb{Q}/\mathbb{Z},U(1)) = \hat{\mathbb{Z}}$, where $\hat{\mathbb{Z}}$ is the group of profinite integers, and it is the inverse limit of $\mathbb{Z}_m$ (see for instance~\cite{Lenstra:ProfiniteGroups}). In practice, an element $\hat{k} \in \hat{\mathbb{Z}}$ can be described as a collection of its values (its images in $\mathbb{Z}_N$) modulo $N$ for all integer $N$. Let $x =\frac{p}{q}, y=\frac{r}{s} , z=\frac{t}{u}$ be elements in $\mathbb{Q}/\mathbb{Z}$ such that $\gcd(p,q) = \gcd(r,s) = \gcd(t,u) =  1$ and $0 \leq p < q$, $0 \leq r < s$, $0 \leq t < u$ with $p,r,t \in \mathbb{Z}$ and $q, s, u \in \mathbb{Z}\setminus \{0\}$. We denote by $M$ the least common multiple of $q, s$ and $u$. 
The 3-cocycle in $H^3(\mathbb{Q}/\mathbb{Z},U(1))$ represented by $\hat{k}$ can be explicitly written as the phase
\begin{equation} \label{eq:QZ3cocycle}
    \omega_{\hat{k}} (x, y, z) = \exp \left(2 \pi i \frac{(\hat{k} \,\ \text{mod}\,\ M) \ p}{q} \lambda(r,s,t,u)\right) \,,
\end{equation} 
where $\lambda(r,s,t,u) = 1$ if $\frac{r}{s} + \frac{t}{u} > 1$ and $\lambda(r,s,t,u)=0$ otherwise.

From the expression~\eqref{eq:QZ3cocycle}, we see that the $\mathbb{Q}/\mathbb{Z}$ part of $U(1)^\delta$ with its third cohomology group being $\hat{\mathbb{Z}}$ captures the anomalies for all $\mathbb{Z}_m \subset U(1)^\delta$. For the details of the above group theoretical discussions and the computation of $H^3(U(1)^\delta, U(1))\supset H^4(BU(1),\mb{Z})$, see Appendix~\ref{sec:H3delta}.


\textit{Higher dimensional generalizations.} We have explicitly derived the connection between the F-move of topological defects and the anomaly polynomial for the flat gauging of the continuous symmetry group $G$, for the $d=2$ case. Note that if we take the dual graph of the right picture in Figure~\ref{fig:Deformations}, we obtain a 3-simplex $v_0 v_1 v_2 v_3$ whose vertices are $v_i$ $(i=0,\dots,3)$ while the edges correspond to elements of $G$. Such picture can be easily generalized to $d=2k$ spacetime dimensions, where the dual 3-simplex $v_0 v_1 v_2 v_3$ is replaced by a dual $(2k+1)-$simplex $v_0 v_1 \dots v_{2k}$. Taking the dual of $v_0 v_1 \dots v_{2k}$, one obtains the higher-dimensional generalization of F-move, as in Dijkgraaf-Witten theory~\cite{Dijkgraaf:1989pz}.

The Lie algebra cohomology $H^3(\mathfrak{g},\mb{R})$ and group cohomology $H^3(U(1)^\delta,U(1))$ should be straightforwardly generalized to $H^{2k+1}(\mathfrak{g},\mb{R})$ and $H^{2k+1}(U(1)^\delta,U(1))$. We present the explicit generators for $H^{2k+1}(\mathfrak{g},\mb{R})$ for semisimple Lie algebra $\mathfrak{g}$, $H^{2k+1}((U(1)^l)^\delta,U(1))$ for a product of $U(1)$ 0-form symmetry groups, and their correspondence with anomaly polynomials, in Appendix~\ref{sec:higher-dimension}.

\vspace{0.3cm}
\begin{acknowledgments}

We would like to thank Jin Chen, Yuji Tachikawa, Zheyan Wan, Xin Wang, Zhaolong Wang and Yunqin Zheng  for useful discussions. YZ would like to thanks Mikhail Kapranov for helpful explanations about the cohomology of Lie groups made discrete. JT and YNW would like to thank Peng Huanwu Center for Fundamental Theory and Northwest University for hosting the $6^\text{th}$ National Workshop on Fields and Strings where many progress of this project has been made. RL and YNW are supported by National Natural Science Foundation of China under Grant No. 12175004, No. 12422503 and by Young Elite Scientists Sponsorship Program by CAST (2024QNRC001). YNW is also supported by National Natural Science Foundation of China under Grant No. 12247103.  QJ is supported by National Research Foundation of Korea (NRF) Grant No. RS-2024-00405629 and Jang Young-Sil Fellow Program at the Korea Advanced Institute of Science and Technology.
YZ is supported by WPI Initiative, MEXT, Japan at Kavli IPMU, the University of Tokyo and by National Science Foundation of China (NSFC) under Grant No. 12305077.

\end{acknowledgments}


\bibliography{refs.bib}

\onecolumngrid

\appendix

\section{'t Hooft Anomaly of $U(1)$ Global Symmetry and Lie Group Cohomology}
\label{sec:anomaly-U(1)}

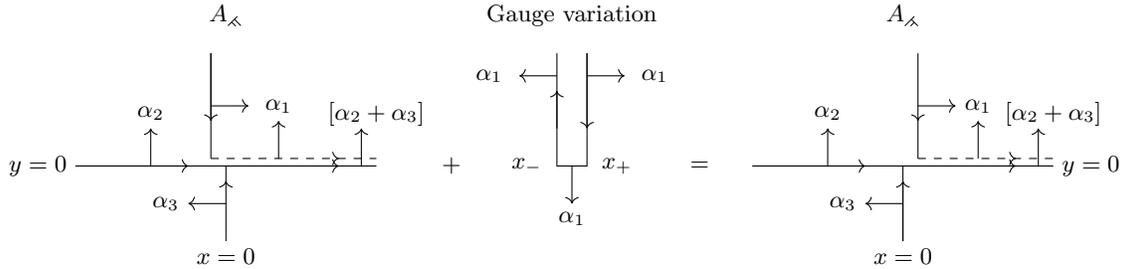
\begin{figure}[h]
	\centering
	\begin{tikzpicture}
		\draw (-2,0) -- (2,0);
		\draw (0,-1) -- (0,0);
		\draw[->] (0,-0.5) -- (-0.5,-0.5);
		\draw[->] (-1,0) -- (-1,0.5);
		\draw (-0.2,0.1) -- (-0.2,1.5);
		\draw[dashed] (-0.2,0.1) -- (2.0,0.1);
		\draw[->] (1.8,0) -- (1.8,0.5);
		\node at (-0.8,-0.5) {$\alpha_3$};
		\node at (-1,0.7) {$\alpha_2$};
		\node at (2.0,0.7) {$[\alpha_2+\alpha_3]$};
		\draw[->] (-0.2,0.8) -- (0.3,0.8);
		\draw[->] (0.7,0.1) -- (0.7,0.6);
		\node at (0.7,0.8) {$\alpha_1$};
		\node at (-2.5,0) {$y = 0$};
		\node at (0,-1.2) {$x = 0$};
		\node at (0,2.0) {$A_{\scalebox{0.3}{\Fbefore}}$};
        \draw[->] (-2,0) -- (-0.5,0);
        \draw[->] (0,0) -- (1.5,0);
        \draw[->] (1.49,0.1) -- (1.5,0.1);
        \draw[->] (-0.2,1.5) -- (-0.2,0.6);
        \draw[->] (0,-0.21) -- (0,-0.2);
		
		\node at (3.0,0) {$+$};

        \draw[->] (4.4,0.5) -- (4.4,1);
        \draw[->] (4.8,1.5) -- (4.8,0.5);
		\draw (4.+0.4,1.5) -- (4.+0.4,0) -- (4.4+0.4,0) -- (4.4+0.4,1.5);
		\draw[->] (4+0.4,1.2) -- (3.5+0.4,1.2);
		\draw[->] (4.4+0.4,1.2) -- (4.9+0.4,1.2);
		\draw[->] (4.2+0.4,0) -- (4.2+0.4,-0.5);
		\node at (4.2+0.4,-0.7) {$\alpha_1$};
		\node at (3.5,1.2) {$\alpha_1$};
		\node at (5.7,1.2) {$\alpha_1$};
		\node at (3.6+0.4,0) {$x_-$};
		\node at (4.8+0.4,0) {$x_+$};
		\node at (4.6,2.0) {Gauge variation};
		
		\node at (6.3,0) {$=$};

        \draw[->] (-2+9,0) -- (-0.5+9,0);
        \draw[->] (0+9,0) -- (1.5+9,0);
        \draw[->] (1.49+9,0.1) -- (1.5+9,0.1);
        \draw[->] (0+9,-1) -- (0+9,-0.2);
        \draw[->] (-0.2+9.4,0.61) -- (-0.2+9.4,0.6);
		\draw (-2+9,0) -- (2+9,0);
		\draw (0+9,-1) -- (0+9,0);
		\draw[->] (0+9,-0.5) -- (-0.5+9,-0.5);
		\draw[->] (-1+9,0) -- (-1+9,0.5);
		\draw[->] (1.8+9,0) -- (1.8+9,0.5);
		\draw (-0.2+9.4,0.1) -- (-0.2+9.4,1.5);
		\draw[dashed] (-0.2+9.4,0.1) -- (-0.2+11.2,0.1);
		\draw[->] (-0.2+9.4,0.8) -- (0.3+9.4,0.8);
		\draw[->] (1.0+9,0.1) -- (1.0+9,0.6);
		\node at (-0.8+9,-0.5) {$\alpha_3$};
		\node at (-1+9,0.7) {$\alpha_2$};
		\node at (2.0+9,0.7) {$[\alpha_2+\alpha_3]$};
		\node at (0.6+9.4,0.8) {$\alpha_1$};
		\node at (-2.5+14,0) {$y = 0$};
		\node at (0+9,-1.2) {$x = 0$};
		\node at (0+9,2.0) {$A_{\scalebox{0.3}{\Fafter}}$};
	\end{tikzpicture}
	\caption{The F-move for $U(1)$ global symmetry.}\label{fig:Fmove_U1-2}
\end{figure}

As we have noted already, the 't Hooft anomaly of $U(1)$ needs extra care since $H^3(\mathfrak{g},\mathbb{R}) = 0$. The F-move is given by the move shown in Figure~\ref{fig:Fmove_U1-2}.
We note that here the key point is that for abelian global symmetry one has to consider the consequence of the periodicity of the parameter. Thus the fusion of $e^{i\alpha_2}$ and $e^{i\alpha_3}$ results in $e^{i[\alpha_2+\alpha_3]}$ rather than naively $e^{i(\alpha_2+\alpha_3)}$. The Poincar\'e dual of the initial web is:
\begin{equation}
	\begin{split}
		A_{\scalebox{0.3}{\Fbefore}} &= i( \alpha_1\delta(x-x_-)H(y) - \alpha_3\delta(x)H(-y) ) dx + i\left( \alpha_1 H(x-x_-) + \alpha_2 H(-x) + [\alpha_2 + \alpha_3] H(x) \right) \delta(y) dy
	\end{split}
\end{equation}
where $0 < \alpha_i < 1$. The field strength is ($[\alpha]\equiv \alpha\ \mathrm{mod}\ 2\pi\mb{Z}$):
\begin{equation}
	F_{\scalebox{0.3}{\Fbefore}} = dA_{\scalebox{0.3}{\Fbefore}} = i([\alpha_2 + \alpha_3] - \alpha_2 - \alpha_3) \delta(x)\delta(y)dx\wedge dy \,.
\end{equation}
For the gauge variation we have:
\begin{equation}\label{eq:shift_A}
	A_{\scalebox{0.3}{\Fafter}} - A_{\scalebox{0.3}{\Fbefore}} = i\alpha_1 \left( -\delta(x-x_-)H(y)dx - B(x;x_-,x_+)\delta(y)dy + \delta(x-x_+)H(y)dx \right) = d\left( - i\alpha_1 B(x;x_-,x_+)H(y) \right)=d\epsilon\,.
\end{equation}
Therefore, we have:
\begin{equation}\label{eq:gauge_param}
	\epsilon = - i\alpha_1 B(x;x_-,x_+)H(y)\,.
\end{equation}
One can compute the anomaly to be:
\begin{equation}\label{eq:anomaly_abelian}
	\begin{split}
		\mathcal{A}[A_{\scalebox{0.3}{\Fbefore}};\epsilon] =\frac{1}{(2\pi)^2}\int_{M_2} \epsilon dA_{\scalebox{0.3}{\Fbefore}} &= \frac{1}{(2\pi)^2}\int_{M_2} \alpha_1  (\alpha_2 + \alpha_3 - [\alpha_2+\alpha_3]) B(x;x_-,x_+)H(y) \delta(x)\delta(y)dx\wedge dy \\
		&= \frac{1}{(2\pi)^2}\alpha_1 \left( \alpha_2 + \alpha_3 - [\alpha_2 + \alpha_3] \right) \,.
	\end{split}
\end{equation}
One immediately recognizes that the above result generalizes the result for discrete global symmetry in~\cite{Moore:1988qv}.

To see that in this case the anomaly is described by the Lie group cohomology $H^3(U(1)^\delta,U(1))$ rather than by the Lie algebra cohomology (which is actually trivial), we recall that the coboundary operator $\delta$ for Lie group cohomology is defined as:
\begin{equation}
	(\delta f) (g_1,\cdots,g_{n+1}) = f(g_2,\cdots,g_{n+1}) \times \prod_{i}  f(g_1,\cdots,g_ig_{i+1},\cdots,g_{n+1})^{(-1)^i} \times f(g_1,\cdots,g_{n})
\end{equation}
For $f(e^{i \alpha_1}, e^{i \alpha_2}, e^{i \alpha_3}) = e^{i\alpha_1 \left( \alpha_2 + \alpha_3 - [\alpha_2 + \alpha_3] \right)/(2\pi)}$ we have:
\begin{equation}
	\begin{split}
		f(e^{i \alpha_2}, e^{i \alpha_3}, e^{ i \alpha_4}) &= e^{i\alpha_2 \left( \alpha_3 + \alpha_4 - [\alpha_3 + \alpha_4] \right)/(2\pi)}\,, \\
		f(e^{i (\alpha_1 + \alpha_2)}, e^{i \alpha_3}, e^{\alpha_4}) &= e^{i[\alpha_1 + \alpha_2]\left( \alpha_3 + \alpha_4 - [\alpha_3 + \alpha_4] \right)/(2\pi)}\,, \\
		f(e^{i\alpha_1}, e^{i (\alpha_2 + \alpha_3)}, e^{ i \alpha_4}) &= e^{i\alpha_1 \left( [\alpha_2 + \alpha_3] + \alpha_4 - [[\alpha_2 +\alpha_3] + \alpha_4] \right)/(2\pi)}\,, \\
		f(e^{i \alpha_1}, e^{i \alpha_2}, e^{i (\alpha_3 + \alpha_4)}) &= e^{i\alpha_1 \left( \alpha_2 + [\alpha_3 + \alpha_4] - [\alpha_2 + [\alpha_3 + \alpha_4]] \right)/(2\pi)}\,, \\
		f(e^{i \alpha_1}, e^{i \alpha_2}, e^{i \alpha_3}) &= e^{i\alpha_1 \left( \alpha_2 + \alpha_3 - [\alpha_2 +\alpha_3] \right)/(2\pi)}\,,
	\end{split}
\end{equation}
We define $n_{ij} := \alpha_i + \alpha_j - [\alpha_i + \alpha_j] \in \{0,2\pi\}$. Hence we have:
\begin{equation}
	\begin{split}
		&\ (\delta f) (e^{i \alpha_1}, e^{i \alpha_2}, e^{ i \alpha_3}, e^{i \alpha_4}) \\
        =&\ \frac{f(e^{i \alpha_2},e^{i \alpha_3},e^{i \alpha_4}) f(e^{i\alpha_1}, e^{i (\alpha_2 + \alpha_3)}, e^{i \alpha_4}) f(e^{i \alpha_1}, e^{i \alpha_2}, e^{i \alpha_3})}{f(e^{ i (\alpha_1 + \alpha_2)}, e^{ i \alpha_3}, e^{ i \alpha_4}) f(e^{ i \alpha_1}, e^{ i \alpha_2}, e^{i (\alpha_3 + \alpha_4)})} \\
		=&\ e^{i ( \alpha_2 n_{34} - (\alpha_1 + \alpha_2 - n_{12})n_{34} + \alpha_1(\alpha_2 + \alpha_3 + \alpha_4 - n_{23} - [\alpha_2 + \alpha_3 + \alpha_4]) - \alpha_1(\alpha_2 + \alpha_3 + \alpha_4 - n_{34} - [\alpha_2 + \alpha_3 + \alpha_4]) + \alpha_1 n_{23} )/(2\pi)} \\
		=&\ e^{i n_{12} n_{34}/(2\pi)} = 1\,.
	\end{split}
\end{equation}
Therefore we have proved that $f(\alpha_1,\alpha_2,\alpha_3) = e^{i\alpha_1 \left( \alpha_2 + \alpha_3 - [\alpha_2 + \alpha_3] \right)/(2\pi)}$ is a cocycle with respect to $\delta$. Since $f$ both and $f^k$ are not $\delta$-exact~\cite{Kawagoe:2021gqi, Jia:2025vrj}, we have proved that the anomaly $e^{2\pi i\mathcal{A}[A_{\scalebox{0.3}{\Fbefore}};\Lambda]} = e^{i \alpha_1\left( \alpha_2 + \alpha_3 - [\alpha_2 + \alpha_3] \right)/(2\pi)}$ is in the subgroup $H^4(BU(1),\mb{Z})\cong\mb{Z}\subset H^3(U(1)^\delta,U(1))$.

\section{Group and Lie algebra cohomology in higher dimensions}
\label{sec:higher-dimension}

We also present the representatives for group and Lie algebra cohomologies with degree $n>3$, which are relevant for anomalies in $(n-1)$ spacetime dimensions.

For abelian group $U(1)$, the representative for the subgroup of the $U(1)$ group cohomology with discrete topology $\mb{Z}\subset H^{2k+1}(U(1)^\delta,U(1))$ is given by
\be
f(e^{i\alpha_1},e^{i\alpha_2},\dots,e^{i\alpha_{2k+1}})=e^{ (2\pi)^{-k}i\alpha_1\prod_{i=1}^k(\alpha_{2i}+\alpha_{2i+1}-[\alpha_{2i}+\alpha_{2i+1}])}\,.
\ee

One can explicitly check the closedness condition
\be
\ba
(\delta f)(e^{i\alpha_1},\dots,e^{i\alpha_{2k+2}})&=f(e^{i\alpha_2},\dots,e^{i\alpha_{2k+2}})\cdot\left(\prod_{i=1}^{2k+1} f(e^{i\alpha_1},\dots,e^{i[\alpha_i+\alpha_{i+1}]},\dots,e^{i\alpha_{2k+1}})^{(-1)^i}\right)\cdot f(e^{i\alpha_1},\dots,e^{i\alpha_{2k+1}})\cr
&=\exp\left(i(2\pi)^{-k} \prod_{i=1}^{k+1}(\alpha_{2i-1}+\alpha_{2i}-[\alpha_{2i-1}+\alpha_{2i}])\right)\cr
&=1\,.
\ea
\ee


We also discuss the group cohomology representatives for mixed anomalies. For the mixed anomaly between $U(1)_1\times U(1)_2$ in 2D, described by the gauge invariant 4-form anomaly polynomial $(2\pi)^{-2}F^{(1)}\wedge F^{(2)}$, it corresponds to the 3-cocycle generator $f\in \mb{Z}\subset H^3(U(1)_1^\delta\times U(1)_2^\delta,U(1))$:
\be
f((e^{i\alpha_1},e^{i\beta_1}),(e^{i\alpha_2},e^{i\beta_2}),(e^{i\alpha_3},e^{i\beta_3}))=\exp(i\alpha_1(\beta_2+\beta_3-[\beta_2+\beta_3])/(2\pi))\,.
\ee
Here $e^{i\alpha_m}\in U(1)_1$ and $e^{i\beta_m}\in U(1)_2$ $(m=1,2,3)$.

More generally in $2k$-spacetime dimensions, for a mixed anomaly between $U(1)_1\times\dots\times U(1)_l$, described by a gauge invariant $(2k+2)$-form anomaly polynomial $(2\pi)^{-k-1}(F^{(1)})^{m_1}\dots (F^{(l)})^{m_l}$, $m_1+\dots+m_l=k+1$, the proposed group cohomology generator $f\in \mb{Z}\subset H^{2k+1}(U(1)_1\times\dots\times U(1)_l,U(1))$ is
\be
\begin{aligned}
&f((e^{i\alpha_1^{(1)}},\dots,e^{i\alpha_1^{(l)}}),\dots,(e^{i\alpha_{2k+1}^{(1)}},\dots,e^{i\alpha_{2k+1}^{(l)}}))\cr
=&\exp(i(2\pi)^{-k} \alpha_1^{(1)}\left(\prod_{m=2}^{m_1}(\alpha_{2m-2}^{(1)}+\alpha_{2m-1}^{(1)}-[\alpha_{2m-2}^{(1)}+\alpha_{2m-1}^{(1)}])\right)\cr
&\cdot\prod_{j=2}^l\prod_{m=1}^{m_l}(\alpha_{2m_1+\dots+2m_{j-1}+2m-2}^{(j)}+\alpha_{2m_1+\dots+2m_{j-1}+2m-1}^{(j)}-[\alpha_{2m_1+\dots+2m_{j-1}+2m-2}^{(j)}+\alpha_{2m_1+\dots+2m_{j-1}+2m-1}^{(j)}]))\,.
\end{aligned}
\ee
Again $e^{i\alpha_m^{(j)}}\in U(1)_j$, $(m=1,\dots,2k+1)$.


\section{Anomalies, cohomology groups and secondary invariants.}
\label{sec:anomaly-coh}

\paragraph{F-move and group cohomologies.} In the case of a finite group $G$, the F-move induces a local transformation of the topological network, with distinct configurations differing by a phase. For 2D bosonic theory, this phase factor is classified by the group cohomology $H^3(BG,\mathbb{R}/\mathbb{Z})$ (we treat $U(1) = \mathbb{R}/\mathbb{Z}$ as an additive group), while in the fermionic case the phase factor is determined by a triple $(\mu,\nu,\alpha) \in Z^1(BG,\mathbb{Z}_2) \times Z^2(BG,\mathbb{Z}_2) \times C^3(BG,\mathbb{R}/\mathbb{Z})$ such that $\delta \alpha= (-1)^{\nu^2}$. The distinction lies in whether the theory depends on spin structure. We will only consider the bosonic case in the continuous group setup. Another crucial feature of a finite group is that the algebraically defined cohomology group $H^\bullet(G,A)$ coincides with the topologically defined cohomology group $H^\bullet(BG,A)$ of the classifying space $BG$.

For a Lie group $G$, both $H^\bullet(BG,A)$ and ``$H^\bullet(G,A)$'' can be defined. The former is defined the same way as in the finite group case by considering the classifying space. However, the latter have various versions. For example, it can be defined in terms of multivariable functions on the group; one can then restrict attention to the continuous functions with respect to the standard topology of the group $G$, this group is usually written as $H^\bullet_c(G,A)$ with the subscript indicating the continuousness~\cite{Stasheff1978}. One can take $G^\delta$ as the group $G$ with discrete topology, in such a way we can define $H^\bullet_c(G^\delta,A)$, which is the analog of $H^\bullet(G,A)$ in the finite group case. However, both $H^\bullet_c(G,A)$ and $H^\bullet_c(G^\delta,A)$ differ from $H^\bullet(BG,A)$. For instance, if $G$ is compact connected and simply connected, $H^n_c(G,\mathbb{R}) = H^n_c(G,\mathbb{R}/\mathbb{Z}) = 0$ for $n >0$.  

As any flat $G$-bundle is a $G^\delta$-bundle, we focus on discussions of $G^\delta$ in the case of flat gauging. The aim is to show that there is a class $[\Omega] \in H^\bullet_c(G^\delta,\mathbb{R}/\mathbb{Z})$ that serves as an analog of the Chern--Simons invariant and it is the anomaly.

\paragraph{Classifying space and secondary invariant.} Given the moduli space of flat connection is $\text{Hom}(\pi_1(M), G)/G$ one can verify~\cite{morita2001geometry} that the classifying space $BG^\delta$ for flat $G$-bundle is the following Eilenberg-Maclane space
\begin{equation}
    BG^\delta \cong  K(G,1) \,.
\end{equation}
Equipped with discrete topology, $G^\delta$ behaves like a discrete group and $H^\bullet (K(G,1),A) \cong H^\bullet (BG^\delta,A)  \cong H_c^\bullet (G^\delta,A)$, where the formed is topologically defined. Note that one has equivalent definitions of flat $G$-bundle due to Corollary (3.22) of~\cite{Dupont1978}
\begin{enumerate}
    \item $G$-bundle admit flat connection;
    \item $G$-bundle admit a set of constant transition functions;
    \item $G$-bundle with a reduction of structure group to $G^\delta$ via the canonical map (actually the identity map)
    \begin{equation}
       \iota: G^\delta \longrightarrow G \,.
    \end{equation}
\end{enumerate}
It is the second definition that allows one to describe a flat $G$-bundle by topological defect networks. 
The third definition can also be rephrased as, there exists a $G^\delta$-bundle whose image under $\iota_*$ (the nature map between $G^\delta$-bundles and $G$-bundles induced by $\iota$) is the flat $G$-bundle~\cite{Kamber1967}. The canonical map $\iota$ also induces a map at the level of classifying space due to the functoriality of $B-$
\begin{equation} \label{eq:Biota}
    B\iota : BG^\delta \longrightarrow BG \,.
\end{equation}
For any flat bundle given by $\varphi:M \rightarrow BG$ there exists a $G^\delta$-bundle by $\varphi^\delta:M \rightarrow BG^\delta$ such that the following diagram commutes
\begin{equation} \label{eq:comdia1}
\begin{tikzcd}
 & M \arrow[dr, "\varphi"] \arrow[dl, "\varphi^\delta"'] & \\
BG^\delta \arrow[rr, "B\iota"'] & & BG
\end{tikzcd}
\end{equation}

 There is a proposition (proposition 9.1 in \cite{Dupont1978}) stating that the composite map is zero for a general Lie group $G$
\begin{equation} \label{eq:proposition91}
    I(G) \xlongrightarrow{\textrm{Chern--Weil}} H^\bullet(BG,\mathbb{R})\xlongrightarrow{B\iota^*}H^\bullet(BG^\delta,\mathbb{R}) \,,
\end{equation}
where $I(G)$ is the set of invariant polynomials.
This proposition tells us that $f \in I^n(G)$ is always mapped to $0\in H^{2n}(BG^\delta,\mathbb{R})$. 

We now consider the following commutative diagram of long exact sequences
\begin{equation}
\begin{tikzcd} \label{eq:diagramchasing}
\cdots  \arrow[r, "j^*"]  & H^{2n-1}(BG^\delta, \mathbb{R}/\mathbb{Z}) \arrow[r, "\beta"] & H^{2n}(BG^\delta, \mathbb{Z}) \arrow[r, "i^*"] & H^{2n}(BG^\delta, \mathbb{R}) \arrow[r, "j^*"] & \cdots\\
\cdots  \arrow[r, "j^*"]  & H^{2n-1}(BG, \mathbb{R}/\mathbb{Z}) \arrow[r, "\beta"] \arrow[u, "B\iota^*"'] & H^{2n}(BG, \mathbb{Z}) \arrow[r, "i^*"] \arrow[u, "B\iota^*"'] & H^{2n}(BG, \mathbb{R}) \arrow[r, "j^*"] \arrow[u, "B\iota^*"'] & \cdots   \,,
\end{tikzcd}
\end{equation}
where the vertical arrows are just $B\iota^*$ of various coefficients and the horizontal long exact sequences are induced by the standard short exact sequence of coefficients
\begin{equation}
    0 \longrightarrow \mathbb{Z} \xlongrightarrow{i} \mathbb{R} \xlongrightarrow{j} \mathbb{R}/\mathbb{Z} \longrightarrow 0 \,,
\end{equation}
and $\beta$ is the connecting homomorphism, or called Bockstein homomorphism, conventionally. 
\begin{itemize}
    \item If $[f] \in H^{2n}(BG,\mathbb{R}) $ is the image of some integral class $[f]_\mathbb{Z} \in H^{2n}(BG,\mathbb{Z})$, i. e. $i^*([f]_\mathbb{Z}) = [f]$. Because $B\iota^*([f])=0 \in H^{2n}(BG^\delta,\mathbb{R})$ and the commutativity of the right quadrat implies that $i^*(B\iota^*([f]_\mathbb{Z})) = 0 \in H^{2n}(BG^\delta,\mathbb{R})$.
    \item Exactness at the stage of $H^{2n}(BG^\delta,\mathbb{Z})$ in the upper sequence implies that there exists a class 
\begin{equation}
    [Tf] \in H^{2n-1}(BG^\delta, \mathbb{R}/\mathbb{Z}) \,,
\end{equation}
such that $\beta([Tf]) = B\iota^*([f]_\mathbb{Z})$, where $\beta$ is the Bockstein homomorphism of the upper long exact sequence.
\end{itemize}
Summarizing the above discussion, we found a canonical map 
\begin{equation}
\begin{split} \label{eq:cano1}
   \kappa^\delta:  H^{2n}(BG,\mathbb{Z}) &\longrightarrow H^{2n-1}(BG^\delta, \mathbb{R}/\mathbb{Z}) \\
   [f]_\mathbb{Z}&\longmapsto [Tf] \,.
\end{split} 
\end{equation}

Note that if we replace $BG^\delta$ by $M$, and using a flat $G$-bundle $\varphi$ get zero map $(\textrm{Chern--Weil} \circ \varphi^*=0)$ (now the composite map is zero because of the vanishing of the curvature 2-form)
\begin{equation} \label{eq:modfiproposition91}
    I(G) \xlongrightarrow{\textrm{Chern--Weil}} H^\bullet(BG,\mathbb{R})\xlongrightarrow{\varphi*}H^\bullet(M,\mathbb{R}) \,,
\end{equation}then this would lead to a class $CS \in H^{2n-1}(M,\mathbb{\mathbb{R}/\mathbb{Z}})$, namely Chern--Simons invariant constructed by Chern and Simons explicitly and we just demonstrated how one can obtain it alternatively via diagram chasing. Hence, fixing a flat bundle $\varphi$, we have another canonical map 
\begin{equation}
\begin{split}  \label{eq:cano2}
   \kappa:  H^{2n}(M,\mathbb{Z}) &\longrightarrow H^{2n-1}(M, \mathbb{R}/\mathbb{Z}) \\
   (c[f])_\mathbb{Z}&\longmapsto CS \,,
\end{split} 
\end{equation}
where $c([f])_\mathbb{Z}$ is the integral characteristic class of $[f]_\mathbb{Z}$ pulled back to $M$, i. e. $(c[f])_\mathbb{Z} = \varphi^*([f]_\mathbb{Z})$. 

Combining \eqref{eq:comdia1}, \eqref{eq:cano1} and \eqref{eq:cano2}   all together, we arrived at the commutative diagram 
\begin{equation}
\begin{tikzcd}
H^{2n}(BG,\mathbb{Z}) \arrow[r, "\kappa^\delta"] \arrow[d,"\varphi^*"'] & H^{2n-1}(BG^\delta, \mathbb{R}/\mathbb{Z}) \arrow[d,"(\varphi^\delta)^*"] \\
H^{2n}(M,\mathbb{Z}) \arrow[r,"\kappa"'] & H^{2n-1}(M, \mathbb{R}/\mathbb{Z}) \,,
\end{tikzcd}
\end{equation}
in which the commutativity follows from the functoriality of $B-$ and we have
\begin{equation} \label{eq:OmegaCS}
    (\varphi^\delta)^*([Tf]) = CS.
\end{equation}
We can apply this to our anomaly discussions. Let $n=2$ and for instance take $G=SU(2)$ and $c([f]_\mathbb{Z}) = c_2$ the second Chern-class. We now denote  $[Tf]$ simply by $[\Omega]$ and obtain straightforwardly $(\varphi^\delta)^*(\Omega) = \kappa(c_2) = CS$.

\paragraph{Transgression and Homotopy fiber.}

Given a fibration 
\begin{equation}
    F \longrightarrow E \xlongrightarrow{p} M\,,
\end{equation}
the \emph{trangression/suspension}~\cite{Borel1955TopologyLieGroups} operation allows one to relate cohomology classes of $M$ to cohomology classes of $F$ with degree shifted by $1$. Mathematically speaking, the details rely on the \emph{Leray-Serre} spectral sequence. Here we list some steps relevant to our discussion
\begin{itemize}
    \item One starts from a cohomology class $[\alpha] \in H^{p+1}(M)$ and assumes that its pullback in $E$ trivializes $[p^*(\alpha)] = 0 \in H^{p+1}(E)$;
    \item Then there exists a $p$-cochain, denoted as $\eta \in C^p(E)$ such that $\delta\eta = p^*(\alpha)$; 
    \item The restriction of $\eta$ to the fiber $F$ is called $\overline{T\alpha}$ and $\delta(\overline{T\alpha})= 0$, one thus obtains a cohomology class $[\overline{T\alpha}] \in H^p(F)$ and $[\alpha]$ is called the \emph{transgression} of $[\overline{T\alpha}]$.
\end{itemize}

A primary example is the universal $G$-bundle 
\begin{equation}
    G \longrightarrow EG \xlongrightarrow{p} BG\,,
\end{equation}
where the cohomology of $EG$ is trivial since it is contractible. Hence we have the \emph{suspension} map by the above process
\begin{equation} \label{eq:susp}
    \tau : H^n(BG,A) \longrightarrow H^{n-1}_{\rm singular}(G,A) \,,
\end{equation}
where $A$ is an arbitrary coefficient and $x \in \text{Im}(\tau) \subset H^{n-1}_{\rm singular}(G,A)$ is called \emph{universal transgressive}. In the case of $A=\mathbb{Z}$ as coefficient and $n=4$, the map $\tau$ is in general not surjective, but for compact, connected, simply connected and semi-simple Lie groups it is an isomorphism (in particular for each compact, connected, simply connected simple factor).

Another important fibration in our consideration can be introduced as follows.

Following the standard construction of \emph{homotopy fiber}, one can write the map $\iota$ into a fibration 
\begin{equation}
    \overline{G} \longrightarrow E_\iota \xlongrightarrow{\iota} G \,,
\end{equation}
where $E_\iota$ is given as 
\begin{equation}
    E_\iota  = \{(g, \gamma) \in G^\delta \times G^{I}| \gamma(0)= \iota(g)\} \subset G^\delta \times G^{I} \,,
\end{equation}
and it is homotopy equivalent to $G^\delta$ and we sometimes just write $G^\delta$ instead. $G^{I}$ is the mapping space of continuous maps from the interval $I=[0,1]$ to $G$, it is a topological space endowed with the compact-open topology.
The homotopy fiber $\bar{G}$ is by construction defined as 
\begin{equation}
    \overline{G} = \{(g, \gamma) \in G^\delta \times G^{I}| \gamma(0)= \iota(g), \gamma(1) = \iota(1_{G^\delta}) \} \subset G^\delta \times G^{I} \,.
\end{equation}

 The homotopy fiber construction also extends to $B\iota$ and we have the following fibration 
\begin{equation} \label{eq:BGfib}
     B\bar{G} \longrightarrow BG^\delta \xlongrightarrow{B\iota} BG \,.
\end{equation}
We can now apply the suspension operation to the fibration \eqref{eq:BGfib} and use the proposition \eqref{eq:proposition91}. Pick $f \in I^k(G)$ such that $[f] \in H^{2n}(BG,\mathbb{R})$, then its determines a class 
\begin{equation}
    [\overline{Tf}] \in H^{2n-1}(B\overline{G}, \mathbb{R})\,,
\end{equation}
that transgresses to $[f]$.

\section{On $H^3(U(1)^\delta,U(1))$}
\label{sec:H3delta}

A mathematically proper definition of the ``$N \longrightarrow \infty$'' limit of $\mathbb{Z}_N$ is via the \emph{direct limit} (see~\cite{Putrov:2022pua} for a discussion in the context of physics applications). That is, build a direct system by the inclusion $\mathbb{Z}_n \rightarrow \mathbb{Z}_m$ whenever $n$ divides $m$ and take the limit of this system
\begin{equation}
    \varinjlim \mathbb Z_n = \mathbb{Q}/\mathbb{Z},
\end{equation}
an explicit description of the limit is by taking the union of all cyclilc subgroups $\mathbb{Z}_n \subset U(1)^\delta$, i.e.  
\begin{equation}
    \mathbb{Q}/\mathbb{Z} = \bigcup_n \,\mathbb{Z}_n\,.
\end{equation}
The group $\mathbb{Q}/\mathbb{Z}$ captures all roots of unity in $U(1)^\delta$ and it is the torsion subgroup of $U(1)^\delta$. Since $U(1)^\delta \cong \mathbb{R}^\delta/\mathbb{Z}$, we have 
\begin{equation}
    U(1)^\delta \cong \mathbb{R}^\delta/\mathbb{Q} \oplus \mathbb{Q}/\mathbb{Z}\,.
\end{equation}
The above isomorphism is due to the fact that the $\mathbb{Q}/\mathbb{Z}$ is divisible and hence the following short exact sequence splits (although not canonical and depends on a choice of basis of $\mathbb{R}^\delta$)
\begin{equation}
    0 \longrightarrow \mathbb{Q}/\mathbb{Z} \longrightarrow \mathbb{R}^\delta/\mathbb{Z} \longrightarrow \mathbb{R}^\delta/\mathbb{Q} \longrightarrow 0 \,.
\end{equation} 
Admitting the axiom of choice, $\mathbb{R}^\delta$ is regarded as infinite dimensional $\mathbb{Q}$-vector spaces with an uncountable basis. In the same vein, the divisible torsion-free group $\mathbb{R}^\delta/\mathbb{Q}$ is also a $\mathbb{Q}$-vector spaces with a uncountable basis. Let us denote the uncountable index set of a basis of $\mathbb{R}^\delta/\mathbb{Q}$ by $I$ then $\mathbb{R}^\delta/\mathbb{Q}= \oplus^I \mathbb{Q}$.
We will also omit the superscript $\delta$ for brevity, as no confusion will arise in what follows.  
We denote the above groups by $A=\mathbb{R}^\delta/\mathbb{Q}$, $B=\mathbb{Q}/\mathbb{Z}$ and $G=\mathbb{R}^\delta/\mathbb{Z}$. At the level of classifying space, we are dealing with $K(G,1) = K(A,1) \times K(B,1)$. We can apply the universal coefficient theorem 
\begin{equation}
        0 \longrightarrow  \text{Ext}_\mathbb{Z}(H_{2}(G,\mathbb{Z}),U(1))\longrightarrow  H^3(G,U(1)) \cong H^3(K(G,1),U(1)) \longrightarrow \text{Hom}_\mathbb{Z}( H_3(G,\mathbb{Z}),U(1))\longrightarrow 0 \,,
\end{equation}
which yields $H^3(G,U(1)) \cong \text{Hom}_\mathbb{Z}( H_3(G,\mathbb{Z}),U(1))$ since $\text{Ext}_\mathbb{Z}(H_{2}(G,\mathbb{Z}),U(1))$ vanishes for the divisible group $U(1)$. As $G= A \oplus B$, we have for $H_3(G,\mathbb{Z})$ the Künneth formula~\cite{HiltonStammbach1971}
\begin{equation} \label{eq:KuennethFormula}
       0 \longrightarrow  \bigoplus_{p+q=3} H_p(A,\mathbb{Z}) \otimes H_q(B,\mathbb{Z}) \longrightarrow  H_3(G,\mathbb{Z}) \longrightarrow \bigoplus_{p+q=2}\text{Tor}( H_p(A,\mathbb{Z}),H_q(B,\mathbb{Z}))\longrightarrow 0 \,,
\end{equation}
and the sequence splits by an unnatural splitting.  
Now recall the standard result (which can be derived using the techniques in Section 3.F of~\cite{hatcher2002algebraic})
\begin{align} \label{eq:QZhomology}
H_q(\mathbb{Q}/\mathbb{Z},\mathbb{Z})=\left\{\begin{array}{ll}
\mathbb{Z} & \quad (q =0 )  \\
\mathbb{Q}/\mathbb{Z} & \quad (q  \quad\text{odd})  \\
0 &\quad (q\quad \text{even}\quad \text{and} \quad q >0 )
\end{array}\right.\,.
\end{align}
As for $H_\bullet(A,\mathbb{Z})$, we can think of it as the homology of the Eilenberg-MacLane space $K(A,1)$. Then $H_0(A,\mathbb{Z}) = H_0(K(A,1),\mathbb{Z}) =\mathbb{Z}$ as Eilenberg-MacLane spaces are connected and $H_1(A,\mathbb{Z}) = H_1(K(A,1),\mathbb{Z})= A$ which equals the abelianization of $\pi_1(K(A,1))= A$ for $A$ is abelian.  

From the torsion product part we need to compute $\text{Tor}( H_0(A,\mathbb{Z}),H_2(B,\mathbb{Z}))$, $\text{Tor}( H_1(A,\mathbb{Z}),H_1(B,\mathbb{Z}))$ and $\text{Tor}( H_2(A,\mathbb{Z}),H_0(B,\mathbb{Z}))$. By inserting the above results and using the properties of the $\text{Tor}$ functor we conclude 
\begin{equation}
    \bigoplus_{p+q=2}\text{Tor}( H_p(A,\mathbb{Z}),H_q(B,\mathbb{Z})) = 0\,.
\end{equation}

Now we can consider the tensor product part of~\eqref{eq:KuennethFormula}, for this we need two extra ingredients $H_2(A,\mathbb{Z})$ and $H_3(A,\mathbb{Z})$ which appear in $\mathbb{Q}/\mathbb{Z} \otimes H_2(A,\mathbb{Z})$ and $\mathbb{Z} \otimes H_3(A,\mathbb{Z}) = H_3(A,\mathbb{Z})$. Given $A = \oplus^I \mathbb{Q}$, $K(A,1)$ can be thought as filtered homotopy colimit of $K(\mathbb{Q}^n, 1)$ over finite $n$ and one can use Künneth formula to show that $H_p(K(\mathbb{Q}^n, 1),\mathbb{Z})$ is a $\mathbb{Q}$-vector space  for all $p > 0$. This is also true for $H_p(K(A, 1),\mathbb{Z})$, as filtered colimit of $\mathbb{Q}$-vector spaces are $\mathbb{Q}$-vector spaces~\cite{Wofsey2019_CohomologyOfKReal1}. $H_2(A,\mathbb{Z})$ as a $\mathbb{Q}$-vector spaces is torsion-free and hence 
\begin{equation}
    \mathbb{Q}/\mathbb{Z} \otimes H_2(A,\mathbb{Z}) = 0\,.
\end{equation}
We are left with $H_3(A,\mathbb{Z})$ and $H_0(A,\mathbb{Z}) \otimes H_3(B,\mathbb{Z}) = \mathbb{Z} \otimes \mathbb{Q}/\mathbb{Z} = \mathbb{Q}/\mathbb{Z}$. 

Summarized the above computation we arrive at 
\begin{equation}
    H^3(G,U(1)) = \text{Hom}(H_3(A,\mathbb{Z}) \oplus \mathbb{Q}/\mathbb{Z},U(1)) = H^3(\mathbb{R}^\delta/\mathbb{Q},U(1)) \oplus \hat{\mathbb{Z}} \,,
\end{equation}
where $\hat{\mathbb{Z}}$ is the profinite integers and can be though as the direct product $\Pi_p \,R_p$ where $R_p$ is the ring of $p$-adic integer. Now there is a canonical embedding of  $\mathbb{Z}$ into $\hat{\mathbb{Z}}$ by sending each integer $n$ to the sequence of its residues modulo all positive integers. This map is modeled by embedding the level-$k$ anomaly polynomial (level-$2k$ Chern--Simons invariant in our normalisation) of $U(1)$ into the 3-cocycle of $\hat{\mathbb{Z}}$ part of $H^3(U(1)^\delta,U(1))$.

Note that in the above consideration, $H_3(A,\mathbb{Z}) = H_3(K(A,1),\mathbb{Z})$ remains an unspecified $\mathbb{Q}$-vector space, possibly of uncountable dimension. Together with the cohomology ring of $H^\bullet(K(A,1),\mathbb{Z})$, this lies beyond our current understanding, and we hope it will one day be clarified by mathematicians.

For non-abelian Lie groups, the complexity of this question increases dramatically. It is in general very hard to determine the cohomology group exactly. We will just briefly mention some contents from~\cite{Milnor1983_LieGroupsMadeDiscrete,FriedlanderMislin1984_ClassifyingSpacesCohomology}. Let $G$ be an arbitrary Lie group with finitely many components, then
\begin{itemize}
    \item the map $B\iota^*: H^\bullet(BG,\mathbb{Z}) \rightarrow H^\bullet(BG^\delta,\mathbb{Z})$ induced by the canonical map $B\iota$ from the Eq.~\eqref{eq:Biota} is injective;

    \item \emph{Friedlander--Milnor conjecture.} The canonical map $B\iota$ induces isomorphisms of homology and cohomology with mod $p$ coefficients, or more generally with any finite coefficient group.
\end{itemize}

\section{Anomaly of gauge transformation}
\label{sec:anomaly-gauge-trans}

In this appendix, we will review the gauge anomaly in 2D, noting that the discussion naturally extends to higher dimensions. Suppose we have a 2D theory living on $M_2$ whose anomaly is characterized by the anomaly polynomial $I_4(F)$
    \begin{equation}
        I_4(F) = \frac{k}{2}\textrm{Tr} F\wedge F\,,
    \end{equation}
which is closed and is locally exact
    \begin{equation}
        I_4(F) = d I_3(A)\,,
    \end{equation}
where $I_3(A)\equiv CS_k(A)$ is the local Chern-Simons density. Consider an infinitesimal gauge transformation parametrized by $\Lambda$, the gauge variation of $I_3(A)$ is closed due to
\begin{equation}
    d \delta_{\Lambda} I_3(A) = \delta_{\Lambda} d I_3(A)=\delta_{\Lambda} I_4(F)=0\,,
\end{equation}
and is also locally exact. Then the anomaly of the partition function $Z(A)$ is computed via the descent equation 
    \begin{equation}
        \delta_{\Lambda} I_3(A) = d \mathcal{A}[A;\Lambda]\,,
    \end{equation}
and takes the form
    \begin{equation}\label{anomaly-inf}
        \delta_{\Lambda} \log Z[A] = 2\pi i \int_{M_2} \mathcal{A}[A;\Lambda]\,.
    \end{equation}

We then consider a finite gauge transformation 
    \begin{equation}
        A \rightarrow g^{-1} Ag+ g^{-1}dg\,,
    \end{equation}
parametrized by a group function $g\equiv g(x,y)$ on $M_2$ and is connected to the identity. Let us construct a 3D manifold $M_2 \times [0,1]$ with $t\in[0,1]$ and extend the 2D gauge function $g$ into $g(t)\equiv g(x,y,t)$ as
    \begin{equation}
        g(x,y,0)=e\,,\quad g(x,y,1)=g(x,y)\,.
    \end{equation}
We also extend the gauge field $A$ on $M_2$ along $[0,1]$ according to
    \begin{equation}
        A(t) = g(t)^{-1}A_0 g(t)+g(t)^{-1} d g(t)\,,
    \end{equation}
where $A_0$ is considered as a trivial extension of $A$ on $M_2$ as $A_0(x,y,t)=A(x,y)$. The 2D anomaly of the finite gauge transformation $g$ connected to the identity is then evaluated as
    \begin{equation}\label{sm-anomaly}
        \delta_g \log Z[A]=  2\pi i\int_{[0,1]\times M_2} CS_k[A(t)] \,.
    \end{equation}
First, one can show that the anomaly given above is independent of the extension $g(t)$ with fixed boundary conditions $g(0)=e,g(1)=g$. The reason is that the Chern-Simons integral is invariant under the gauge transformation as long as we fix the boundary condition. Second, if $g(t)$ is small, we have
\begin{equation}
    \int_{[0,1]\times M_2} CS[A(t)]=\int_{[0,1]\times M_2} CS[A(t)] - \int_{[0,1]\times M_2} CS[A_0] = \int_{[0,1]\times M_2} \delta_{g(t)}CS[A]\,,
\end{equation}
where in the middle we use $CS[A_0]=0$ since $A_0$ is a trivial extension. As shown above, the gauge variation of Chern-Simons density is a total derivative
    \begin{equation}
        \delta_{g(t)} CS(A) = d\mathcal{A}[A;\Lambda]\,,
    \end{equation}
where $\Lambda(t)$ is the infinitesimal gauge parameter related by $e^{i\Lambda(t)}=g(t)$. Using the fact $g(0)=e$ and $\Lambda(0)=0$, we recover the infinitesimal version of the gauge anomaly \eqref{anomaly-inf}.

\end{document}